\documentstyle[12pt,aaspp4,flushrt]{article}

\begin{document}
 
\title{Discovery of a New Quadruple Lens HST 1411+5211
\footnote{Based on observations with the NASA/ESA {\it Hubble Space Telescope}
obtained at the Space Telescope Science Institute, which is operated by the
Association of Universities for Research in Astronomy Inc., under NASA contract
NAS 5-26555.}$^,$\footnote{This work used the facilities of the Canadian
Astronomy Data Centre, operated by the National Research Council, Herzberg
Institute of Astrophysics, Dominion Astrophysical Observatory, and partially
funded by the Canadian Space Agency.}}

\author{Philippe Fischer\footnote{Hubble Fellow}$^,$\footnote{Dept. of
Astronomy, 830 Dennison Bldg., University of Michigan, Ann Arbor, MI 48109}}

\author{David Schade\footnote{Dominion Astrophysical Obs., HIA/NRC of
Canada, 5089 West Saanich Rd., RR 5, Victoria, BC, Canada V8X 4M6}}

\author{L. Felipe Barrientos\footnote{Dept. of Astronomy, University of
Toronto, 60 St. George St., Toronto, Canada, M5S 3H8}}

\abstract

Gravitational lensing is an important tool for probing the mass distribution of
galaxies.  In this letter we report the discovery of a new quadruple lens HST
1411+5211 found in archived WFPC2 images of the galaxy cluster
CL140933+5226. If the galaxy is a cluster member then its redshift is
$z=0.46$. The images of the source appear unresolved in the WFC implying that
the source is a quasar. We have modeled the lens as both a single galaxy and a
galaxy plus a cluster. The latter model yields excellent fits to the image
positions along with reasonable parameters for the galaxy and cluster making
HST 1411+5211 a likely gravitational lens.  Determination of the source
redshift and confirmation of the lens redshift would allow us to put strong
constraints on the mass distribution of the lensing galaxy.


\section{Introduction}

Gravitationally lensed systems have a variety of astronomical uses. The most
obvious is that they provide a direct probe of foreground mass distributions
(stars, galaxies, and clusters). Systems which exhibit multiple images of a
single source have an additional application; if the source varies, the system
can be used to measure Hubble's constant (\cite{re64}). Additionally, the
statistics of multiple gravitational lenses can provide a direct and powerful
test of cosmic structure formation theories (\cite{tu84}).

To date eight quadruple lenses have been found, seven with observed lensing
galaxies. Detailed modeling of these systems, has revealed the need for
substantial external shear in two of the lenses and misaligment between the
major axes of the light and matter in one of the lenses (\cite{wi97}).
Increasing the database of known quadruple lens will improve our knowledge of
galaxy mass distributions, allowing us to put tight constraints on the
mass-to-light ratios and hence the dark matter content in the inner few
kiloparsecs. If sufficent lenses can be found over a large redshift range
evolution of galaxy mass-to-light ratios could also be measured.

In this letter we report the discovery of a new quadruple lens system found in
archived WFPC2 images.  The primary lens, which is clearly identified, is a
galaxy located near and quite likely in a cluster at $z = 0.46$.  In \S
\ref{observations} we discuss the observations and the optical appearance of
the lensing galaxy. In \S \ref{lens} we give a brief review of gravitational
lensing physics and apply one single-lens and two double-lens models to the
data. Throughout the paper, we use $\Omega_0=1$ and use $h=$H$_0/100$.

\section{Observations} \label{observations}

Observations of the cluster CL140933+5226 (3c295) were obtained on 25 July 1994
with the Wide-Field Planetary Camera on HST (images were obtained from the
archive). The total integration time was 12600 seconds in 6 individual
exposures using the F702W filter. The 6 images (archive IDs are u2c40a01t,
u2c40a03t, u2c40a05t, u2c40a02t, u2c40a04t, u2c40a06t.c0h) were registered and
combined using an average with cosmic ray (positive) outlier rejection.

The images were reduced and analysed as part of a larger program to perform
two-dimensional surface photometry to derive structural and photometric
parameters for elliptical galaxies in clusters over a wide range of redshift
(\cite{sc97}). Each fitted galaxy was inspected for goodness of fit and the
presence of irregular residuals. The lensed background images were not
recognized until the best-fit galaxy model was subtracted (see Figure 1).

The lensing galaxy is Number 162 from Table 6 of \cite{dr92} but has no
spectroscopically measured redshift. There is a published {\it photometric}
redshift (\cite{th94}) of $z=0.598 \pm 0.11$ (galaxy 103) based on nine narrow
bandpasses spanning 4500\AA\ - 8000\AA. This redshift is suspect for several
reasons. The first is that of the approximately 140 photometric redshifts
determined in the paper galaxy 103 has the largest quoted uncertainty, two to
three times higher than than most of the other galaxies. Secondly, the galaxy
is classified as type Scd based on its spectral energy distribution (SED),
however, the luminosity distribution in the HST images is inconsistent with a
disk system and is an excellent fit to a deVaucouleurs $r^{1/4}$ profile (see
below). These two points seem to imply that the SED of this galaxy is unusual
and since the quoted redshift differs by less than 1.5$\sigma$ from the cluster
redshift ($z=0.46$) we assume the galaxy is a cluster member at redshift
$z=0.46$ for the rest of this paper. 

The position of the lensing galaxy determined using the STSDAS `metric' is
(RA,Dec) = (14:11:19.6,52:11:29) (J2000).  The best-fit observed parameters for
the elliptical galaxy are half-light radius $R_e=0.61\pm 0.03\arcsec$,
ellipticity $\epsilon_L=0.27\pm 0.03$, position angle $38.3\pm 1$ degrees and
total magnitude F702W(AB)$=20.78\pm 0.05$. Using the assumed redshift of
$z=0.46$ and the \cite{co80} spectral energy distributions (convolved with a
filter response function and converted to $M_\nu$ as in \cite{li95}) we can
convert to r(gunn). We find r(gunn) = F702W(AB)$+0.51$ for a present-day
elliptical spectral energy distribution. Thus our F702W(AB) magnitude
corresponds to r(gunn)$=21.29 \pm0.05$ which agrees with the previously
measured value of \cite{dr92} of $r=21.20$. Their value of $R_e=0.6\arcsec$ for
the half-light radius also agrees well with our value of $R_e=0.61\arcsec$
although we find that an exponential disk (their profile classification)
provides a very poor fit to this galaxy while a deVaucouleurs $r^{1/4}$ profile
fits very well.

Assuming a redshift of $z=0.46$ leads to M$_B$(AB) $=-19.32+5\log(h)$, R$_e=2.1
h^{-1}$ kpc, and restframe B central surface brightness $\mu_{0,B}$(AB)
$=13.57$ mag arcsec$^{-2}$. The adopted k-corrections are for a present-day
elliptical galaxy and because the observed band corresponds to 4780\AA\ in the
galaxy's rest-frame (not far from $B$) the effect of a reasonable amount of
color evolution will be very small ($< 0.05$ mag).

An estimate of the expected velocity dispersion can be made by assuming that
this galaxy lies on the fundamental plane. Surface brightness evolution of
elliptical galaxies (\cite{sc97}) with redshift will effect this estimate
significantly.  To compare with \cite{jo96} we first transform the surface
brightness to the Gunn $r$-band. A present-day elliptical has (B-r)(AB)$=1.18$
yielding a central surface brightness $\mu_0$(r,AB) $=12.39$ and
r(gunn)=r(AB)+0.21 yielding $\mu_0(r)=12.60$. For a deVaucouleurs $r^{1/4}$ law
the surface brightness at $R_e$ is $\mu_e=\mu_0+8.33$ and the mean surface
brightness within $R_e$ is $<\mu_e>=\mu_e-1.39$ so that $<\mu_e>_r=19.54$
mag/arcsec$^2$. \cite{jo96} give, for the Gunn $r$-band, $\log
<I_e>=-0.4(<\mu_e>-26.4)$ to yield a value of $\log <I_e>=2.74$. The
fundamental plane we are using is defined at $z\sim 0$. If we assume a surface
brightness evolution law such as that found in \cite{sc97} then we should
correct the surface brightness by $\sim 0.5$ mag to bring it to its present-day
(non-evolved) value. This yields $\log <I_e>=2.54$.

Using our measured value of $\log R_e=0.31 h^{-1}$ kpc allows us to read an
estimate from Figure 1(b) of \cite{jo96}. If we ignore the evolution correction
then we obtain $1.24\log \sigma -0.82 \log <I_e>=0.6$ which is equivalent to
$\sigma=195\pm 20$ km sec$^{-1}$.  On the other hand the evolution corrected
surface brightness yields $\sigma=150\pm 15$ km sec$^{-1}$.  The quoted errors
are estimated from the scatter in Figure 1(b) of \cite{jo96}.

In Table \ref{tablepos} we show the positions of the four lensed background
images relative to the lens galaxy (0,0). The positions and uncertainties were
determined by measuring the locations on each of the six exposures and taking
the mean and standard error. Also given are the F702(AB) magnitudes for each
image. The uncertainties reflect only photon noise and not systematics due to
galaxy subtraction errors (which may be significant for image D). The faintest
image (D) is detected at a signal-to-noise of 12. The galaxy is located in WFC2
and the four images appear unresolved. Therefore, the source is likely to be a
quasar.



%

\section{Lens Models} \label{lens}

\subsection{Gravitational Lens Theory}

In order to test the lensing hypothesis we have attempted to model the system
with simple lens models. To briefly review, for a given source position the
lensed image positions are given by the extrema of the virtual time delay
surface (\cite{bl86}, \cite{ko91}):

\begin{equation} \label{timedelay}
\Delta{t} = {1 \over 2}(x-u)^2+ {1\over 2}(y-v)^2 - \phi(x,y),
\end{equation}

\noindent 
where ($x,y$) and ($u,v$) are angular coordinates in the image plane and source
plane, respectively. The two-dimensional effective potential is:

\begin{equation}
\phi = {2D_{LS}\psi \over D_{OL}D_{OS}c^2},
\end{equation}

\noindent
where $\psi$ is the two-dimensional projected potential and $D_{LS}, D_{OL}$,
and $D_{OS}$ are the angular diameter distances between the lens and source,
observer and lens, and observer and source, respectively. The extremes of Eqn
\ref{timedelay} occur when:

\begin{equation} 
u = x - {\partial{\phi} \over \partial{x}} {\rm ~~and~~ } v = y
- {\partial{\phi} \over \partial{y}},
\end{equation}

\noindent and the image magnification is given by:

\begin{equation}
M^{-1} = \left(1-{\partial^2\phi \over \partial x^2}\right)\left(1
-{\partial^2\phi \over
\partial y^2}\right)-\left({\partial^2\phi \over 
\partial x\partial y}\right)^2.
\end{equation}

\subsection{Fitting the Data}

The fitting procedure involved varying lens model parameters and source
position until the best fit to the measured image positions was obtained.  In
our model fitting we decided to only use the lensed image positions and not the
relative magnifications due to the high probability of microlensing by stars in
the lens galaxy (\cite{wi95}).  The goodnesss of fit was determined from:

\begin{equation}
\chi^2=\sum_{i=1}^{n_c}{(M_i-O_i)^2 \over err_i^2},
\end{equation}

\noindent
where $n_c$ is the number of constraints, $M_i$ is the predicted value for the
observable $O_i$, and $err_i$ is the corresponding uncertainty. In our case
$n_c = 8$ (the four image positions). The number of degrees of freedom is DOF =
$n_c - n_{par}$ where $n_{par}$ is the number of model parameters (including 2
for the source position).

The first model we tried was a single lens centered on the galaxy. We assumed
an edge-on, axisymmetric, oblate, singular isothermal ellipsoid mass
distribution. The projected mass distribution is (\cite{ke97}):


\begin{equation}
\Sigma(R,\theta)={v_{G0}^2(1-\epsilon_M^2) \over
2GR\sqrt{\epsilon_M^2+1}\sqrt{\epsilon_M^2+1-2\epsilon_M\cos2(\theta-
\theta_\epsilon)}}
\end{equation}

\noindent where $\epsilon_M$ is the ellipticity (one minus the axis ratio) of
the projected mass distribution, $\theta_M$ is the position angle of the
projected major axis and $v_{G0}^2$ is related to the mean line-of-sight
velocity dispersion by:

\begin{equation}
{<v^2_{los}> \over v_{G0}^2} = {\tan^{-1}a \over a} 
\end{equation}

\noindent where $a=\epsilon_M(2-\epsilon_M)/(1-\epsilon_M)^2$.  The
corresponding two-dimensional effective potential is (\cite{ka93}):

\begin{eqnarray}
\phi_1=&4\pi{v_{G0}^2D_{LS} \over c^2D_{OS}}{(1-\epsilon_M^2) \over
\sqrt{\epsilon_M^3+\epsilon_M}} r\left\{\cos(\theta-\theta_\epsilon)
\sin^{-1}\left[{2\sqrt{\epsilon_M} \over 1+\epsilon_M}
\cos(\theta-\theta_\epsilon)\right]\right. \nonumber \\
&\left.+\sin(\theta-\theta_\epsilon)
\sinh^{-1}\left[{2\sqrt{\epsilon_M}\over 1-\epsilon_M}
\sin(\theta-\theta_\epsilon) \right]\right\}
\end{eqnarray}

where $r=(x^2+y^2)^{0.5}$ and $\theta=\tan^{-1}(y/x)$ are the polar coordinates
of the angular position on the sky with respect to the galaxy center.  The
parameters $v_{G0}$, $\epsilon_M$, $\theta_\epsilon$ and the source position
are varied in the fitting procedure until the lens reproduces the observed
lensed image positions as closely as possible. The best fit parameters along
with $\chi^2$ are shown in Table \ref{tablelens}. This model is a very poor fit
to the data, the value of $\chi^2$ / DOF is unacceptably high, and the implied
ellipticity for the galaxy is also higher than one would expect from its
optical appearance.

The lensing galaxy is a member of a cluster whose center is located roughly
perpendicular to the line joining images A and C. The second model we tried
consisted of two singular isothermal spheres, one centered on the galaxy and
one centered on the dominant cluster galaxy (offset is 8.66\arcsec\ E and
39.76\arcsec\ N from the primarly lensing galaxy) representing the cluster mass
distribution. The effective potential in this case is:

\begin{equation}
\phi_{2} = 4\pi{D_{LS} \over c^2D_{OS}}(v_{G0}^2r+v_{C0}^2r^{\prime}).
\end{equation}

\noindent This model produces a better but still very poor fit to the data
($\chi^2$ / DOF = 21.19 / 4). 

The galaxy appears elliptical in the WFPC2 image motivating us to try a model
consisting of a singular isothermal sphere for the cluster and an isothermal
elliptical potential for the galaxy.

\begin{equation}
\phi_{3} = \phi_1 + 4\pi{v_{C0}^2D_{LS} \over c^2D_{OS}}r^{\prime}.
\end{equation}

\noindent
This model provided a very good fit to the image positions ($\chi^2$ / DOF =
1.58 / 2). Additionally, the inferred values for the line-of-sight velocity
dispersions, $v_{los} = 175$ and 135 km s$^{-1}$ for $z_s = 1$ and 3,
respectively, are close to the predictions from the fundamental plane
relationship (see \S\ref{observations}). The cluster velocity dispersion is
also within the range seen for other similar clusters. However, the inferred
position angle of the mass distribution major axis differs by about $65\deg$
from the major axis of the luminosity distribution. Attempts to force the mass
to be oriented identically to the light produced $\chi^2$ values larger than
model 2 (spherical galaxy) for non-zero ellipticity. Despite this, we conclude
that the success of the simple double-lens models in fitting the image
positions implies a high probability that HST 1411.3+5211 is a lensed system.

In fact the lens mass model must be more complex. For example, the individual
galaxies in the cluster will cause deviations from a smooth mass
distribution. However, until the redshift of the source is determined (and the
redshift of the lens is verified) and more constraints are found (perhaps from
VLBI imaging) there is little point in adding complexity to the lens model.

\section{Conclusion}

In this letter we report the discovery of a new quadruple lens HST 1411+5211
discovered in deep WFPC2 images of a galaxy cluster at $z=0.46$. The redshift
of the primary lensing galaxy is currently unknown but it is quite likely a
member of the cluster. The images of the source appear unresolved in the WFC
implying that the source is a quasar. We have successfully applied simple lens
models to the data. Acceptable fits require both an elliptical galaxy mass
distribution and an external cluster. Simpler models consisting of only a
galaxy or spherical galaxy plus cluster do not provide viable fits to the data.
The model galaxy velocity dispersion is consistent with the inferred value
based on the fundamental plane relationship, however, the mass distribution
appears to be misaligned from the light distribution by approximately $65\deg$.
We conclude that HST 1411+5211 is a very strong candidate for a gravitationally
lensed system. Determination of the source redshift and confirmation of the
lens redshift would allow us to put strong constraints on the mass properties
of the lensing galaxy.

Support for this work was provided by NASA through grant \# HF-01069.01-94A.

{}

\begin{deluxetable}{crrr}
\tablewidth{0pt} 
\tablecaption{Image Positions \tablenotemark{1}} 
\tablehead{ \colhead{Image} & \colhead{RA} & \colhead{Dec} & F702(AB) \nl
\colhead{} & \colhead{(arcsec)} & \colhead{(arcsec)} & \colhead(mag.)}
\startdata
A &$ 1.115\pm0.003$  & $-0.399\pm0.004$ & $24.96\pm0.03$ \nl
B &$ 0.120\pm0.025$  & $ 0.612\pm0.016$ & $25.00\pm0.03$ \nl
C &$-1.125\pm0.007$  & $ 0.019\pm0.005$ & $24.92\pm0.03$ \nl
D &$-0.261\pm0.005$  & $-0.703\pm0.007$ & $25.95\pm0.10$ \nl

\enddata
\tablenotetext{1}{Galaxy at (0,0)}
\label{tablepos}
\end {deluxetable}


\begin{table}
\plotone{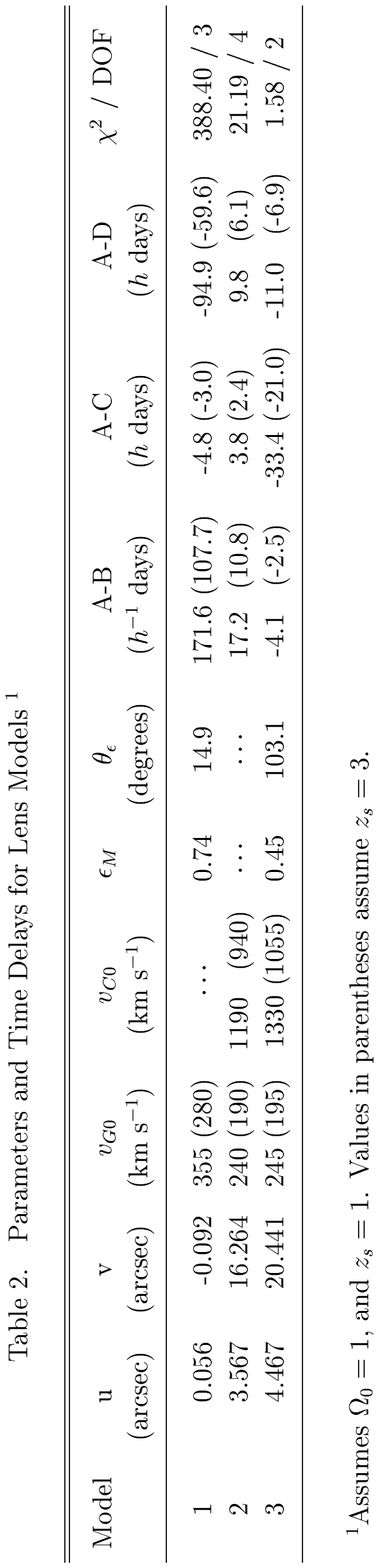}
\label{tablelens}\end{table}

\begin{figure}
\plotone{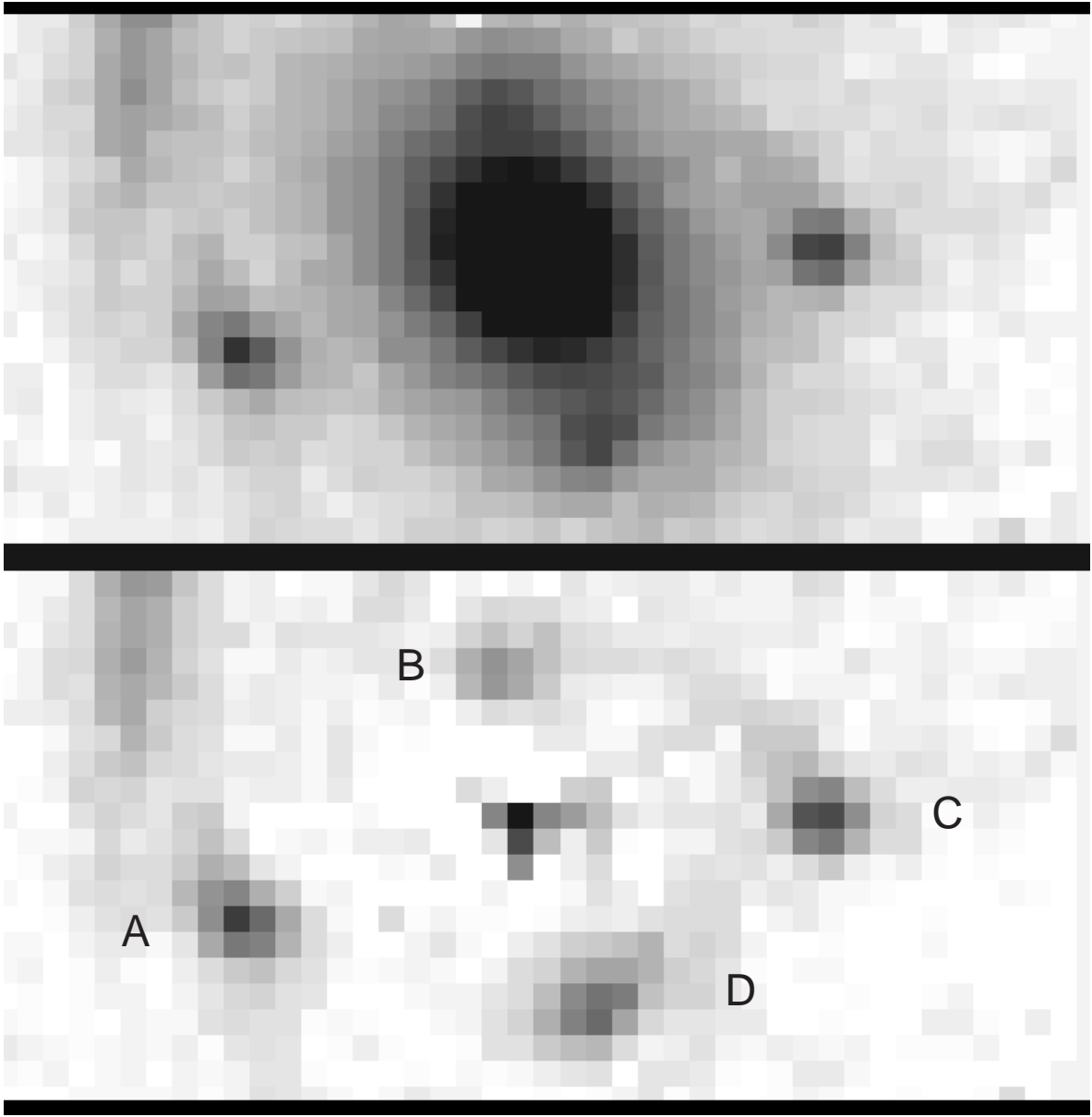}
\caption{The upper panel shows the combined HST image of the regions around the
lensing galaxy. The lower panel shows the same regions with a galaxy model
subtracted. Each panel is $4.3\arcsec \times 2.0\arcsec$. North is up and east
is to the left.}
\end{figure}

\end{document}